\documentclass[aps,prb,twocolumn,showpacs,amsmath,amssymb,superscriptaddress]{revtex4}

\usepackage{graphicx}
\usepackage{hyperref}
\usepackage{float}

\begin{document}

\title{Spin transport in a graphene normal-superconductor junction in the quantum Hall regime}

\date{\today}

\author{Tibor Sekera}
\affiliation{Department of Physics, University of Basel,
Klingelbergstrasse 82, CH-4056 Basel, Switzerland}
%\email[Email address: ] {tibor.sekera@unibas.ch}
\author{Christoph Bruder}
\affiliation{Department of Physics, University of Basel,
Klingelbergstrasse 82, CH-4056 Basel, Switzerland}
\author{Rakesh P. Tiwari}
\affiliation{Department of Physics, McGill University, 3600 rue University, Montreal, Quebec, Canada H3A 2T8}

\date{\today}

\begin{abstract}
  The quantum Hall regime of graphene has many unusual properties. In
  particular, the presence of a Zeeman field opens up a region of
  energy within the zeroth Landau level, where the spin-up and
  spin-down states localized at a single edge propagate in opposite
  directions. We show that when these edge states are coupled to an
  s-wave superconductor, the transport of charge carriers is
  spin-filtered. This spin-filtering effect can be traced back to the
  interplay of specular Andreev reflections and Andreev
  retro-reflections in the presence of a Zeeman field.
\end{abstract}

\pacs{81.05.ue, 73.43.-f, 73.20.At, 74.45.+c
% 81.05.ue : Graphene
% 73.43.-f : Quantum Hall effects
% 73.20.At: Surface states, band structure, electron density of states
% 74.45.+c: Proximity effects; Andreev reflection; SN and SNS junctions
}

\maketitle

\section{Introduction}
Monolayer graphene has remarkable electronic transport properties.
One of them is a peculiar quantum Hall effect, which can be observed
even at room temperature~\cite{novoselov07room}.  Inducing
superconductivity via the proximity effect further enriches these
transport
properties~\cite{heersche07bipolar,jeong11observation,lee15ultimately}.
Recently, a number of experiments have performed conductance
measurements in the quantum Hall regime in monolayer graphene, using
superconducting
electrodes~\cite{rickhaus12quantum,park17propagation,lee17inducing}. Moreover,
coupling the helical edge states within the zeroth Landau level in
graphene to an s-wave superconductor can also give rise to Majorana
bound states~\cite{san-jose15majorana,finocchiaro18topological}.

\begin{figure}
\centering
\includegraphics[width=\columnwidth]{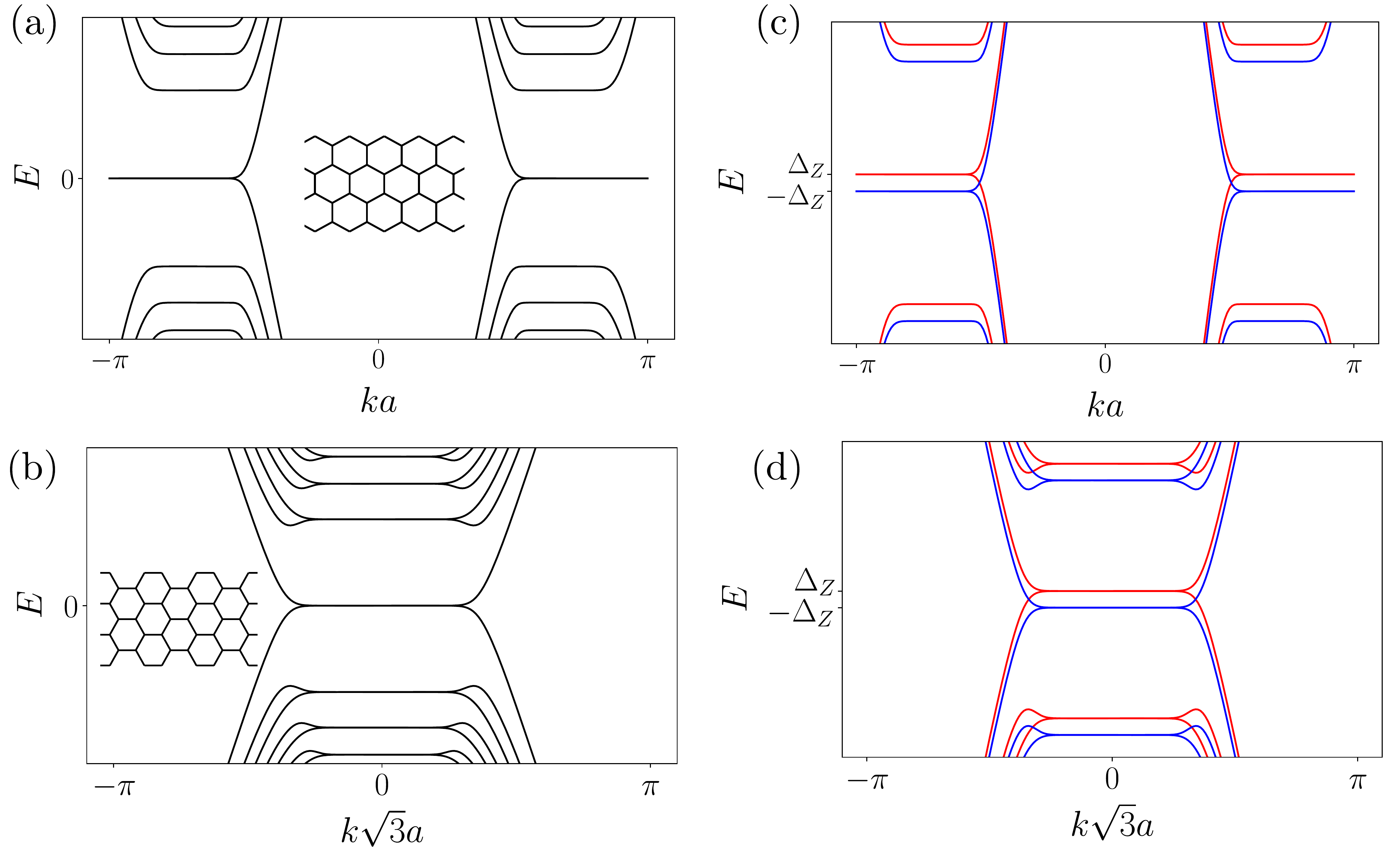}
\caption{Electron band structures of the few lowest Landau
  levels~\cite{abanin06spin}. (a) Band structure of a zigzag ribbon,
  (b) of an armchair ribbon. The ribbons are shown in the insets. (c)
  and (d) The corresponding band structures in the presence of a
  Zeeman field that splits the energies for spin-up (red) and
  spin-down (blue) electrons.}
\label{fig:bands}
\end{figure}

Low-energy excitations in graphene reside in two disconnected regions
in the first Brillouin zone, known as valleys. In the quantum Hall
regime, the energy spectrum has an unconventional Landau level (LL)
structure, where the LL energies are proportional to
$\pm\sqrt{n}$ with integer $n$.  This discrete set of flat LLs develop into dispersive edge states toward the edge of a
sample. In the low-energy approximation, the bulk LL
energies in graphene are given by
\begin{equation}
E^\xi_n = \lambda\hbar\omega_c\sqrt{n}\:, 
\end{equation}
where the valley index $\xi=\pm$ denotes the $K^{\pm}$ valley, and
$\lambda=\pm 1$ labels the conduction and valence band, respectively.
The cyclotron frequency is given by
$\hbar\omega_c=\sqrt{2}\hbar v_F/\ell_B$, where $v_F$ is the Fermi
velocity, $\ell_B=\sqrt{\hbar/(eB)}$ is the magnetic length and
$B=|\mathbf{B}|$ is the absolute value of the applied magnetic
field; $n$ is a nonnegative integer. These bulk LLs are fourfold
degenerate: twofold for the spin and twofold for the valley degree
of freedom. The valley degeneracy is lifted at the edge of the sample,
where the boundary condition for the wavefunction couples the
valleys~\cite{peres06electronic,brey06edge,abanin06spin,akhmerov07detection,tworzydlo07valley}.
Hence the zeroth Landau level (ZLL) splits into two
spin-degenerate bands, one with positive and one with negative
energies, see Figs.~\ref{fig:bands}(a) and (b).

If the spin degeneracy is lifted by, e.g., a Zeeman field, each
of the LLs splits into two with energy difference $2\Delta_Z$, where
$\Delta_Z=\frac{1}{2} g^* \mu_B B$. Here, $g^*$ is the effective
$g$-factor of an electron in graphene and $\mu_B$ is the Bohr
magneton. The energy difference between the spin-up and spin-down bulk
LLs is $2\Delta_Z \approx 2.3\,\text{meV}$ at
$B\sim 10\,\text{T}$ for the interaction-enhanced $g$-factor, $g^*=4$,
see Refs.~\onlinecite{abanin06spin,volkov12interaction}. Close to the edge, the spin
splitting leads to spin-up and spin-down edge states propagating in
opposite directions in the energy region $- \Delta_Z < E < \Delta_Z$,
see Figs.~\ref{fig:bands}(c) and~\ref{fig:bands}(d).  Such a system can be used as a
spin filter. In Ref.~\onlinecite{abanin06spin}, the authors propose
a four-terminal device where the spin-filtering effect can be achieved
by inducing backscattering between the counterpropagating edge states
locally (using gates) in just one part of the system. The
spin-filtering effect takes place due to the presence of an in-plane
magnetic field and spin-orbit coupling.

Here, we suggest a different mechanism for the spin-filtering effect.
We couple the edge states to an s-wave superconductor with a critical
field high enough such that superconductivity and the quantum Hall effect
coexist, and consider only subgap transport. The Andreev-reflected
hole can have the same or the opposite direction of propagation as the
electron impinging on the interface with the superconductor at energy
$E$. Which case is realized depends on the nature of the Andreev
reflection in graphene that can be a retro-reflection for $E<E_F$ or a
specular reflection for $E>E_F$, see
Ref.~\onlinecite{beenakker06specular}.  Hence, if an incoming
spin-down electron is specularly reflected while the spin-up electron
is retro-reflected, spin-filtering takes place. We demonstrate this
effect in the three-terminal device shown in
Fig.~\ref{fig:sample}. This is done by employing a tight-binding model
on a honeycomb lattice within the Bogoliubov-De Gennes framework
and taking into account the orbital and spin effect of the magnetic
field. Note that the present work is related to but different from the
idea in Ref.~\onlinecite{greenbaum07pure}. There, the charge component
of a spin-polarized current is filtered away by using specular
Andreev processes in the absence of an external magnetic
field. However, the authors need a ferromagnetic lead to initially
generate the spin-polarized current.

The rest of this article is organized as follows. In
Sec.~\ref{Sec:Model} we describe the setup of a three-terminal device
and introduce its Hamiltonian. The transport coefficients of this
structure are introduced and determined in
Sec.~\ref{Sec:Transport_coefficients}. We discuss and
summarize our results in Sec.~\ref{Sec:Discussion} and conclude in
Sec.~\ref{Sec:Conclusion}.

\section{Model}
\label{Sec:Model}
\begin{figure}
\centering
\includegraphics[width=0.6\columnwidth]{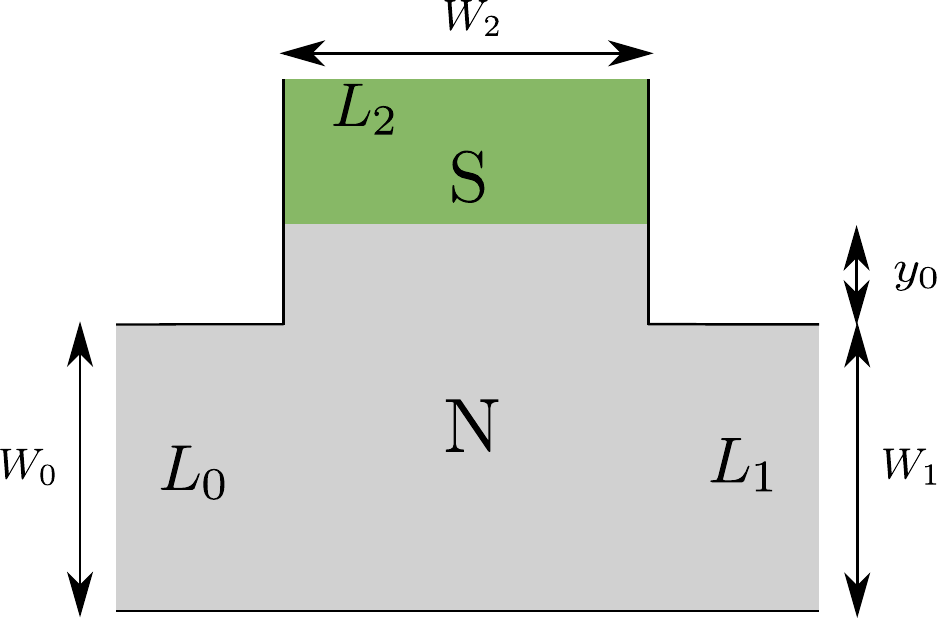}
\caption{A graphene ribbon in the normal state (gray region N) with
  two leads $L_0$ and $L_1$ and a superconducting lead $L_2$ (green
  region S) attached to the top edge. The width of the lead $L_i$ is
  denoted by $W_i$. An external magnetic field smaller than the
  critical field of the superconductor is applied such that the normal
  region is in the quantum Hall regime. The superconductor couples
  electron and hole edge states propagating along the upper edge. We
  assume that a bias voltage $V$ is applied between leads $L_0$ and
  $L_1$.}
\label{fig:sample}
\end{figure}
We investigate spin transport in the three-terminal device shown in
Fig.~\ref{fig:sample}. The underlying honeycomb lattice with lattice
constant $a$ is exposed to a quantizing out-of-plane magnetic
field. The upper edge of the system is coupled to an s-wave
superconductor (S) with a sizable critical field, such that the
quantum Hall effect and superconductivity
coexist~\cite{rickhaus12quantum,park17propagation,lee17inducing}. There
are two normal leads $L_0$ and $L_1$ of widths $W_0$ and $W_1$,
respectively, which serve to probe the spin-resolved transmission
through the scattering region. In the rest of the paper,
$W_0=W_1=W$. The superconducting lead $L_2$ effectively creates a
normal-superconducting interface of length $W_2$ that converts
electrons to holes. The geometry of the system is motivated by a
recent experiment~\cite{park17propagation}.

The tight-binding Hamiltonian of the system can be written as
\begin{equation}
H = H_0 + H_\Delta + H_Z\:,
\label{eq:BdG_Hamiltonian}
\end{equation}
where
\begin{equation}
\begin{aligned}
H_0 =  & \sum_{\langle ij\rangle} \psi^\dagger_i
\left[-te^{i\varphi_{ij}}\frac{1}{2}(\eta_0+\eta_z) \right. \\
& %\phantom{\sum_{\langle ij\rangle} \psi^\dagger_i -} 
+ \left. te^{-i\varphi_{ij}}\frac{1}{2}(\eta_0-\eta_z) \right]\otimes s_0 \psi_j \\ 
&-  E_F\sum_i \psi^\dagger_i (\eta_z\otimes s_0) \psi_i\:, \\
H_\Delta = & \sum_i \Delta_i\psi^\dagger_i (\eta_x\otimes s_0)\psi_i\:, \\
H_Z =      & \sum_i \Delta_{Zi}\psi^\dagger_i (\eta_0\otimes s_z) \psi_i\:.  
\end{aligned}
\end{equation}
The four-spinor field $\psi_i$ is in the standard Nambu basis
$\psi_i=(c_{i\uparrow},c_{i\downarrow},c^\dagger_{i\downarrow},-c^\dagger_{i\uparrow})^T$,
where $\psi^\dagger_i$ creates a particle localized at site $i$ with a
four-component wavefunction
$(\chi_{e\uparrow}(\mathbf{r} -
\mathbf{r}_i),\chi_{e\downarrow}(\mathbf{r} -
\mathbf{r}_i),\chi_{h\uparrow}(\mathbf{r} -
\mathbf{r}_i),-\chi_{h\downarrow}(\mathbf{r} -
\mathbf{r}_i))^T$. Here, the index $es$ $(hs)$ denotes
an electron (hole) with spin $s \in \{ \uparrow, \downarrow \}$.
The two sets of Pauli matrices, $\eta_\nu$ and $s_\nu$ with
$\nu \in \{0,x,y,z\}$, describe the electron-hole and spin degree of
freedom, respectively. Finally, $\sum_{i}$ and
$\sum_{\langle ij \rangle}$ denote sums over all sites and over
nearest neighbors.

The first (second) term in $H_0$ describes the nearest-neighbor
hopping of electrons (holes) in an out-of-plane magnetic field with a
hopping amplitude $-te^{i\varphi_{ij}}$ ($te^{-i\varphi_{ij}}$). The
Peierls phase is given by
\begin{equation}
\varphi_{ij} =  -\frac{2\pi}{\phi_0}B\frac{y_i+y_j}{2}(x_j-x_i)\:,
\end{equation}
where $\phi_0=h/e$ is the magnetic flux quantum and $(x_i, y_i)$ are
the real-space coordinates of site $i$.  The vector potential in the
Landau gauge is chosen to be constant along the $x$-axis,
$\mathbf{A} = (-By,0,0)$.  The third term in $H_0$ represents the
Fermi energy $E_F$ of the system. In undoped graphene, $E_F=0$.

The s-wave superconducting pairing is represented by $H_\Delta$ and
couples an electron with spin $s$ to a hole with spin $s$ on the same
lattice site. The Zeeman field described by $H_Z$ splits each energy level into
two with energy difference $2\Delta_Z$.  For simplicity, we assume the
spatial dependence of the pair potential $\Delta_i$ and of the
magnetic field $B_i$ to be a step function. That is,
$\Delta_i=\Delta(y)$ ($B_i=B(y)$) is assumed to be a non-zero constant
(zero) in the graphene sheet below the superconducting electrode and
zero (a non-zero constant) otherwise. The magnitude of the Zeeman term
has the same spatial dependence as the magnetic field.

In the following, we will calculate the scattering matrix for the
system shown in Fig.~\ref{fig:sample}. All the numerical results for
the conductances and spin polarizations presented below were obtained
using Kwant~\cite{groth14kwant}.

\section{Transport coefficients}
\label{Sec:Transport_coefficients}
\begin{figure}
\centering
\includegraphics[width=0.99\columnwidth]{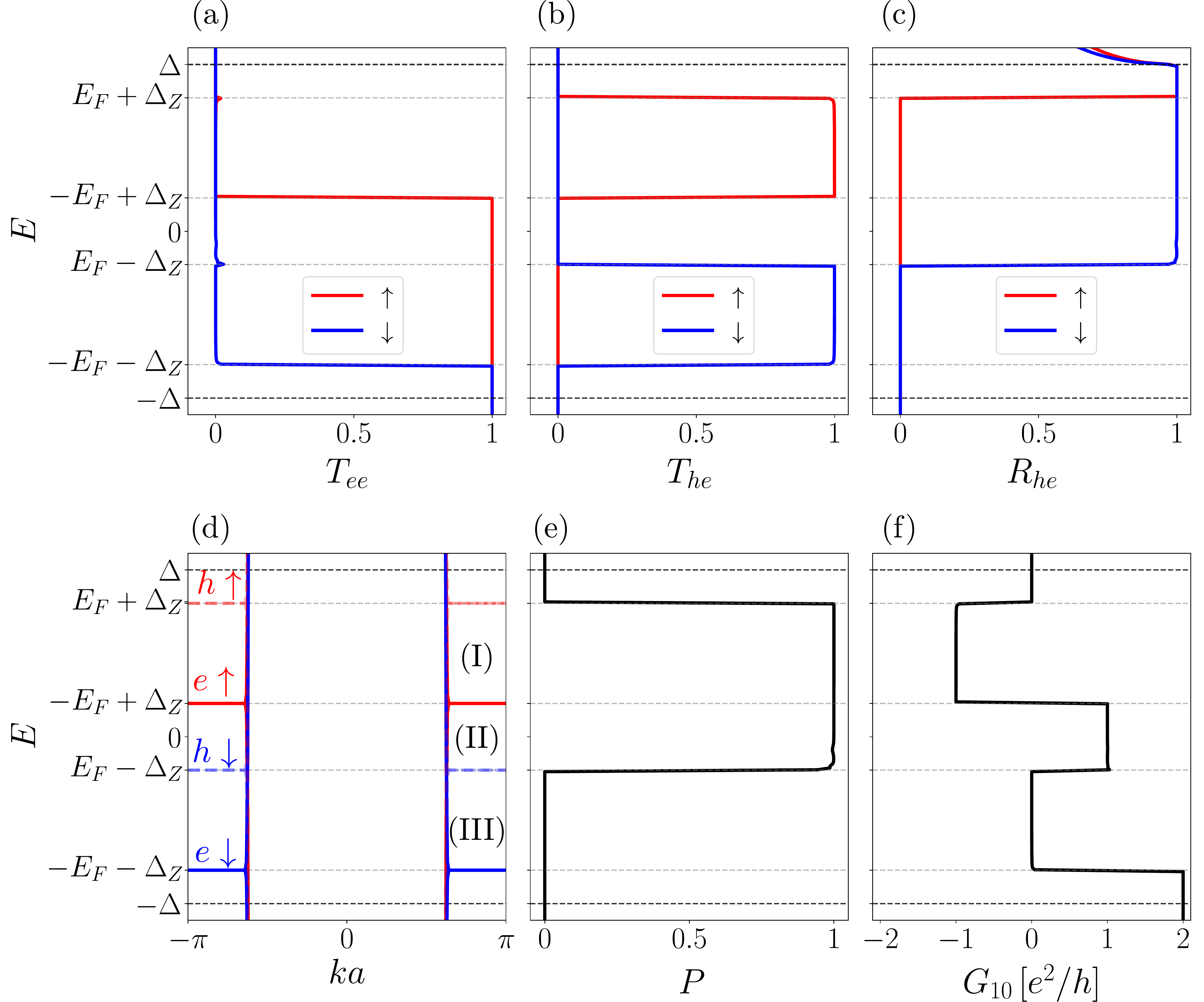}
\caption{On the horizontal axis, we plot in (a)--(c) the transport
  coefficients $T_{ee}$, $T_{he}$, and $R_{he}$ for spin-up (red) and
  spin-down (blue) particles as a function of energy $E$ (vertical
  axis).
  Similarly, (d) shows the band structure of the spin-split zeroth
  Landau level for electrons (full lines) and holes (dashed lines) of
  the normal lead $L_0$, (e) the spin polarization, both as a function
  of $E$. (f) (differential) charge conductance as a function of
  $E=eV$, where $V$ is the bias voltage applied between leads $L_0$ and $L_1$. 
  The thin horizontal dashed lines mark the energies where the edge
  states change the direction of propagation, while the thick ones
  correspond to $|E| = \Delta$. 
  Here, the edge terminations of $L_0$ and $L_1$ are zigzag while the
  edge termination of $L_2$ is armchair. The parameters are
  $\Delta=10\,\text{meV}$, $E_F=0.3\Delta$, $\Delta_Z=0.5\Delta$,
  $B=10\,\text{T}$, $W=600a$, $W_2=510a$, and $y_0=300a$.
}
\label{fig:results}
\end{figure}

In Figs.~\ref{fig:results}(a)--\ref{fig:results}(c) we plot the relevant transport
coefficients in the case when $E_F < \Delta_Z < \Delta$ and the gap
between the ZLL and other LLs is large enough so that only the ZLL
plays a role. Since the Hamiltonian in Eq.~\eqref{eq:BdG_Hamiltonian}
conserves the $z$-projection of the spin, $[H,s_z]=0$, only the
spin-diagonal transport coefficients are shown. The transmission
coefficient for a particle with spin up scattered to a particle with
spin up is shown in red, while blue is used for spin-down
particles. $T_{ee}$ ($T_{he}$) is the probability for an electron from
$L_0$ to be scattered into an electron (a hole) in $L_1$ and $R_{he}$
is the probability for an electron from $L_0$ to be backscattered as a
hole to $L_0$. Because we are in the quantum Hall regime and our
system is wide enough, the probability for an electron from $L_0$ to
be backscattered as an electron is zero ($R_{ee}=0$) and hence not
shown. It can be seen that for energies $|E|<\Delta$, the scattering
matrix is unitary and $T_{ee}+T_{he}+R_{he}=1$ for each spin
projection.

It is interesting to look at the spin polarization of the carriers in
the subgap regime, where $|E|<\Delta$. Since $H$ conserves the spin
projection along the $z$-axis, we define the spin polarization as
\begin{equation}\label{eq:polarization}
P=\frac{T_{e\uparrow,e\uparrow}+T_{h\uparrow,e\uparrow}-T_{e\downarrow,e\downarrow}-T_{h\downarrow,e\downarrow}}{T_{e\uparrow,e\uparrow}+T_{h\uparrow,e\uparrow}+T_{e\downarrow,e\downarrow}+T_{h\downarrow,e\downarrow}}\:,
\end{equation}
where $T_{\alpha' s',\alpha s}$ is the transmission coefficient for a
particle $\alpha$ with spin $s$ in lead $L_0$ to a particle $\alpha'$
with spin $s'$ in lead $L_1$. To avoid numerical artifacts, we set
$P=0$ if the denominator in Eq.~\eqref{eq:polarization} is smaller
than $10^{-3}$, i.e., if almost no particle is transmitted from $L_0$
to $L_1$.
\begin{figure}
\centering
\includegraphics[width=0.99\columnwidth]{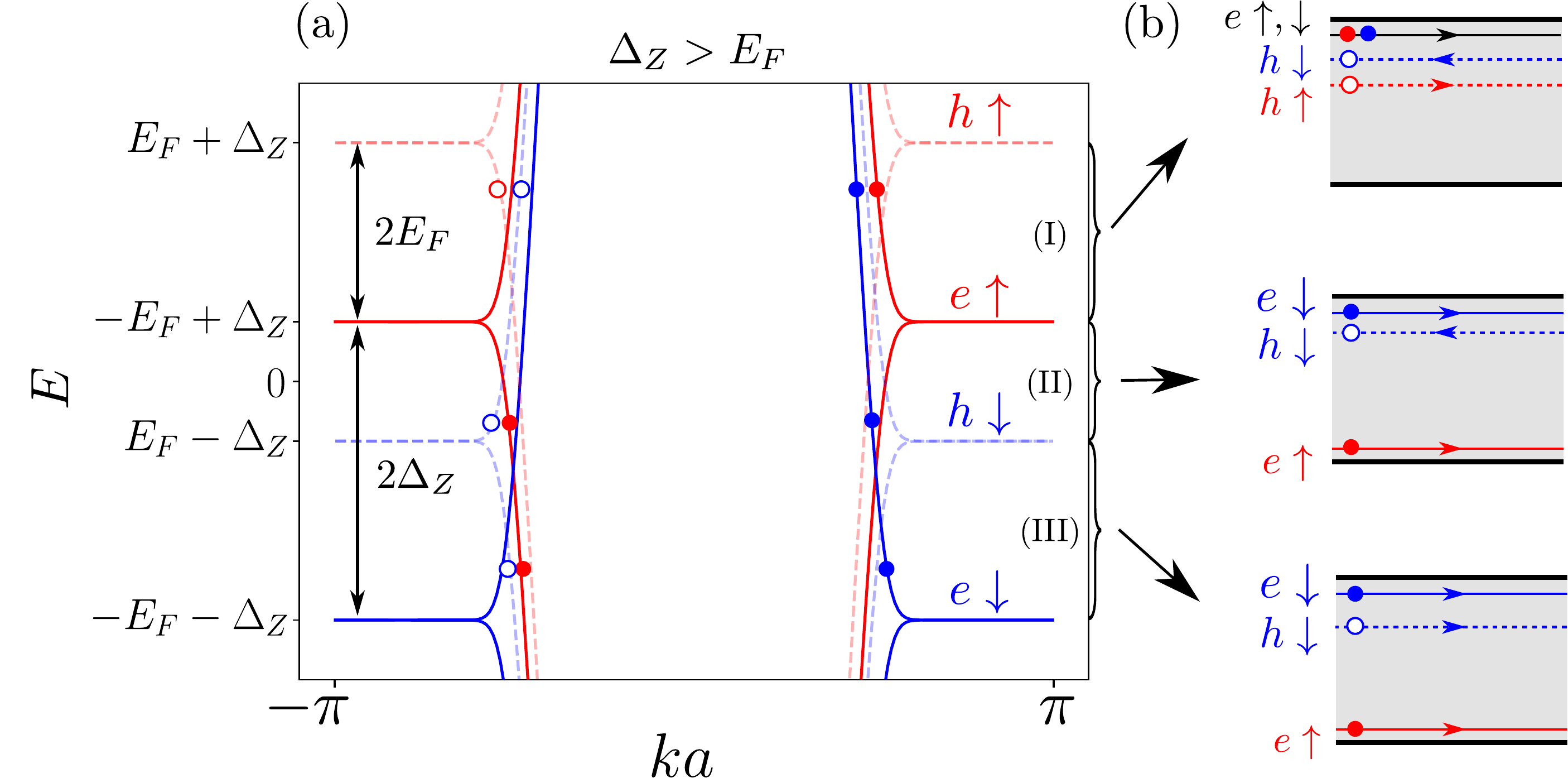}
\caption{
  (a) Band structure of electrons (full lines) and holes (dashed
  lines) of the spin-split zeroth Landau level for spin up (red) and
  spin down (blue) in a graphene zigzag ribbon.  (b) Electron and hole
  edge states in lead $L_0$ and their propagation direction indicated
  by arrows shown for the three energy regions I--III. While there are
  four edge states at each edge, we show only the states relevant for
  the transport in our geometry (see Fig.~\ref{fig:sample}). The
  representative electron (full circle) and hole state (empty circle)
  for spin up (red) and spin down (blue) in each energy region is
  marked in (a) as well as in (b).}
\label{fig:directions}
\end{figure}
The numerically calculated spin polarization is non-zero in the energy
region $E_F-\Delta_Z<E<E_F+\Delta_Z$ and zero otherwise, see
Fig.~\ref{fig:results}(e). This can be understood by looking at the
bandstructure and the propagation direction of the particles along the
edges of the sample as illustrated in Figs.~\ref{fig:directions}(a)
and~\ref{fig:directions}(b), respectively. In the energy region II, a spin-up electron
$e\uparrow$ travels undisturbed along the lower edge into $L_1$,
however a spin-down electron $e\downarrow$ propagating along the upper
edge is backscattered to $L_0$ as a spin-down hole $h\downarrow$
because a superconductor is coupled to the upper edge. This results in
the accumulation of spin-up particles in $L_1$. The situation in the
energy region I is the same for spin-down electrons
$e\downarrow$. However, here a spin-up electron $e\uparrow$ also
travels along the upper edge and encounters the superconductor. Since
an Andreev-reflected spin-up hole $h\uparrow$ has the same propagation
direction as a spin-up electron $e\uparrow$, the particle propagates
along the graphene-superconductor interface via Andreev edge states, and, depending on the
geometry, ends up with a certain probability as a spin-up electron
$e\uparrow$ or spin-up hole $h\uparrow$ in $L_1$. Thus, injecting
spin-unpolarized particles in $L_0$ results in spin-polarized
particles in $L_1$ in the energy region $E_F-\Delta_Z<E<E_F+\Delta_Z$.

We would now like to discuss the (differential) charge
conductance. Here and in the following, we assume the temperature to
be $T=0$. Therefore, the energy $E$ is experimentally given by the
bias voltage, $E=eV$, where $V$ is the potential difference between
the leads $L_0$ and $L_1$. In the presence of hole excitations, the
charge conductance from $L_0$ to $L_1$ is defined as
\begin{equation}
  G_{10}=\frac{e^2}{h} \sum_{s=\uparrow,\downarrow}\left(
    T_{es,es}-T_{hs,es}\right)\:,
\end{equation}
which is shown for our system in Fig.~\ref{fig:results}(f). In the
energy region (I) the carrier ending in $L_1$ is a hole and
$G_{10}=-e^2/h$, while in the region II it is an electron and
$G_{10}=e^2/h$. In the energy region III there is a spin-up electron
$e\uparrow$ along the lower edge and a spin-up hole $h\uparrow$
along the upper edge propagating into $L_1$, which results in zero
charge transfer and $G_{10}=0$. The charge conductance behavior,
however, is not universal and depends on the valley structure of the
edge states~\cite{akhmerov07detection}.

\begin{figure}
\centering
\includegraphics[width=\columnwidth]{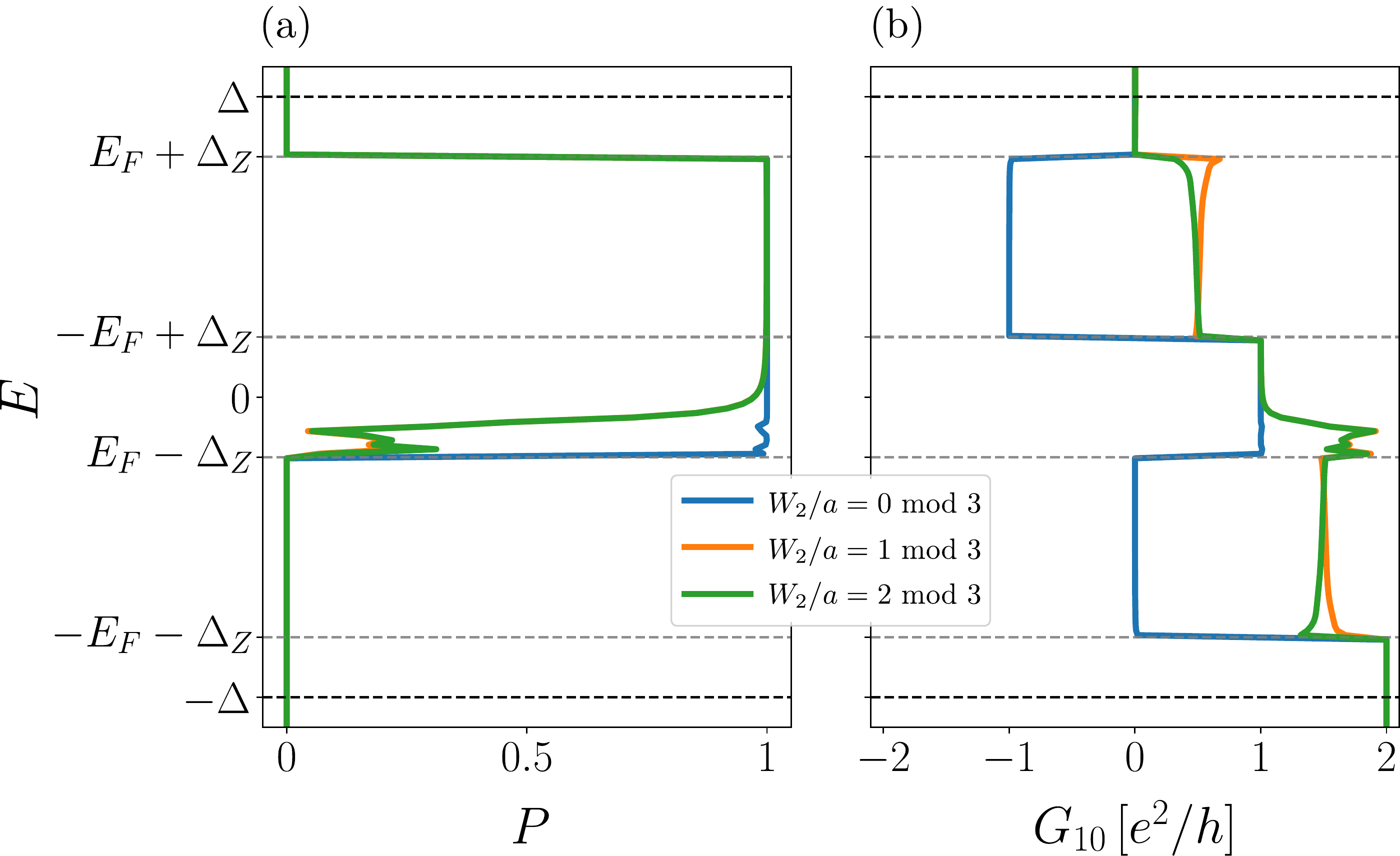}
\caption{(a) Spin polarization and (b) (differential) charge conductance for
  three different interface lengths (values of $W_2$) that correspond to three
  different valley polarizations. The charge conductance depends on
  the angle between the valley isospins, while the spin polarization
  does not (up to the region where $(e\downarrow)$ leaks to $L_1$ due
  to the smaller induced gap.) Here, $\Delta=20\,\text{meV}$,
  $E_F=0.3\Delta$, $\Delta_Z=0.5\Delta$, $B=10\,\text{T}$, 
  $W=600a$, and $y_0=300a$.
}
\label{fig:different_charge_cond}
\end{figure}

\section{Discussion}
\label{Sec:Discussion}
\begin{figure}
\centering
\includegraphics[width=0.7\columnwidth]{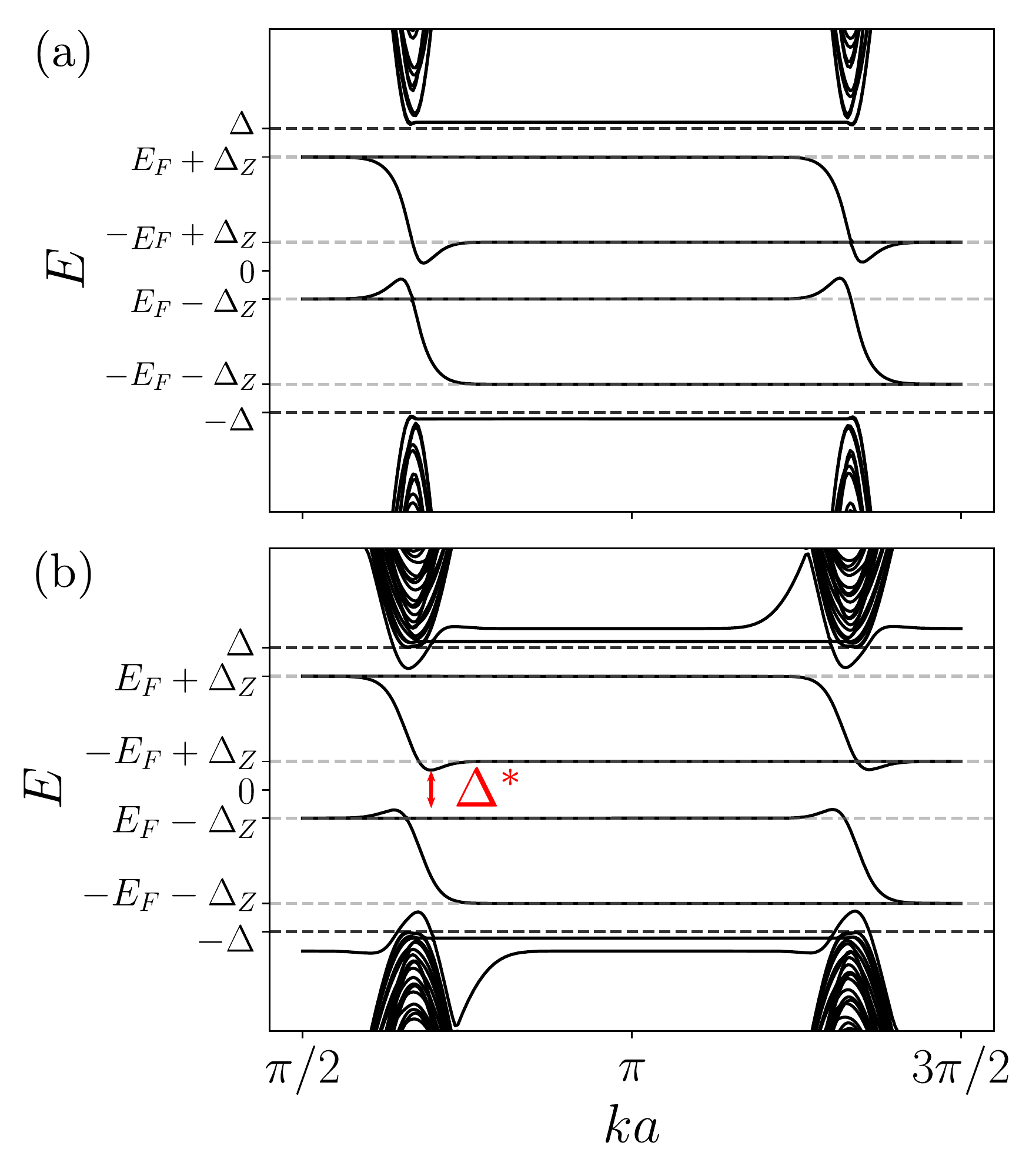}
\caption{(a)--(b) Band structure of excitations along the
  graphene-superconductor interface in a quantizing magnetic field for
  $\Delta/t=0.05$ and $0.1$, respectively. Dispersing states at the
  interface for $|E|<\Delta$ evolve into the flat zeroth Landau level
  (ZLL) away from the interface. The bulk ZLL is split into four:
  electrons and holes are coupled via the superconductor, while the
  spin degeneracy is lifted due to the Zeeman field. The interface
  states are valley-degenerate since the interface is smooth on the
  scale of the lattice constant. When $E_F < \Delta_Z$, the ZLL edge
  states develop an effective band gap $\Delta^*$ (red arrows) due to
  the coupling to a superconductor. $\Delta^*$ increases from (a) to
  (b) with increasing superconducting pair potential $\Delta$. Here,
  $E_F=0.3\Delta$, $\Delta_Z = 0.5\Delta$,
  %$\phi=0.004\phi_0$ which corresponds to $B = 315.6\,\text{T}$
  and the interface is along the zigzag direction. The parameters
  $\Delta$ and magnetic field are chosen to be larger than their
  realistic values to obtain better visibility.}
\label{fig:induced_gap_opening}
\end{figure}

In Fig.~\ref{fig:different_charge_cond} we show the spin polarization
and conductance for three different widths $W_2$ of lead $L_2$ which
is assumed to have armchair edge termination. We see that the spin
polarization is (almost) independent of the interface length $W_2$,
while the charge conductance has a threefold character, depending upon
the total number of hexagons across the width of the armchair ribbon
being a multiple of three, or a multiple of three plus/minus
one~\cite{akhmerov07detection}. Besides that, a set of dips (peaks) in
the spin polarization (conductance) for energies close to
$E_F - \Delta_Z$ can be observed. This feature is due to a spin-down
electron $e\downarrow$ leaking from $L_0$ to $L_1$ through the
interface (without being Andreev-reflected). This can be understood as
follows. Without the superconductor, there are edge states propagating
in opposite directions for opposite spins. When we couple the
superconductor to the upper edge, the electron impinging on the
interface will be reflected as a hole (in the case of non-zero Andreev
reflection probability). However, this hole propagates in the
direction opposite to the electron edge state (for both spin
projections) in this energy region. Hence, the transport along the
interface should be blocked. But if the Andreev reflection probability
is less than one, the electron has a finite chance to leak along the
interface onto the other side. In other words, edge states along the
upper edge contacted to a superconductor develop an effective
gap~\cite{san-jose15majorana} $\Delta^*$ that is smaller than the
naively expected gap $2(\Delta_Z-E_F)$ (for
$(\Delta_Z-E_F)<\Delta$). The bigger the pairing $\Delta$, the higher
the Andreev reflection probability. Thus, on increasing $\Delta$,
$\Delta^*$ approaches $2(\Delta_Z - E_F)$ as shown in
Fig.~\ref{fig:induced_gap_opening}.
\begin{figure}
\centering
\includegraphics[width=\columnwidth]{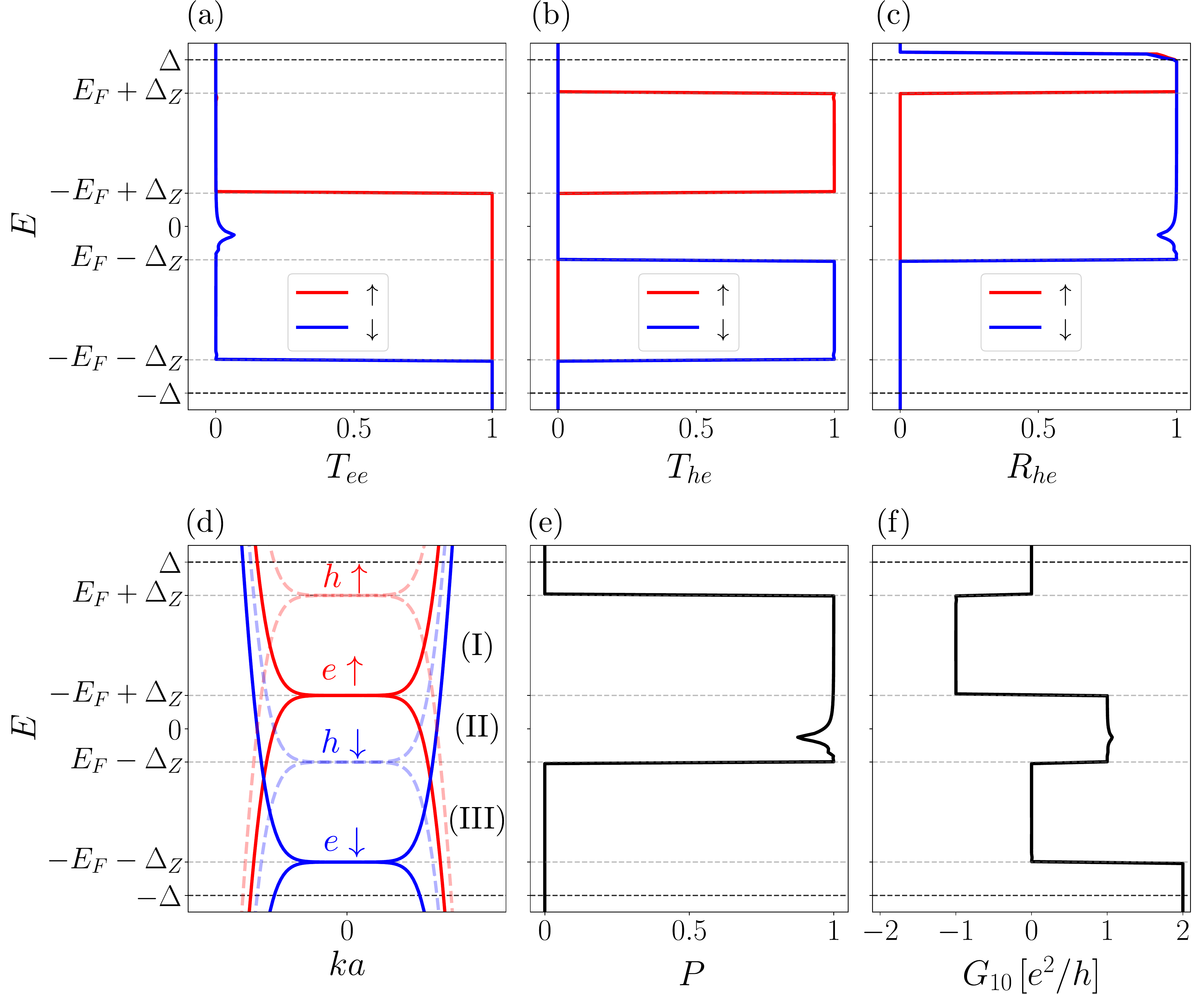}
\caption{Same as Fig.~\ref{fig:results} but with armchair edge
 terminations for $L_0$ and $L_1$ and zigzag edge termination for $L_2$. 
% the parameters for the bandstructure (d): Delta=0.004 EF=0.3*Delta
% Zeeman=0.5*Delta W=300 L=250.0 sw=250.0 y0=150 phi=0.00035 V0=0.0 
}
\label{fig:results_ac}
\end{figure}

The spin-filtering effect for $|E|<\Delta$ is lost once the gate
voltage shifts the Fermi energy such that it exceeds
$\Delta+\Delta_Z$, i.e., the propagation direction of the electron and
hole states is the same within the subgap region.

We obtain similar results if leads $L_0$ and $L_1$ have armchair
orientation and lead $L_2$ has zigzag orientation, see
Fig.~\ref{fig:results_ac}. The spin polarization in
Fig.~\ref{fig:results_ac}(e) is again (nearly) perfect for
$E_F-\Delta_Z<E<E_F+\Delta_Z$. This is expected since, unlike the
valley structure, the spin structure of the ZLL in graphene is
independent of the type of the edge termination. The conductance
profile in Fig.~\ref{fig:results_ac}(f) matches the one in
Fig.~\ref{fig:different_charge_cond}(b) for $W_2/a=0 \text{ mod } 3$,
which is the result of the same valley structure for the states at the
edges of the graphene-superconductor interface for the two cases. The dip in the spin
polarization is present for the same reason as in
Fig.~\ref{fig:different_charge_cond}.

% Let us now mention the expected behaviour in simpler limits.
In the absence of the superconducting proximity effect ($\Delta=0$ in
the graphene sheet), the spin filtering takes place in the energy
region $-E_F-\Delta_Z<E<-E_F+\Delta_Z$ and the hole excitations play
no role. The spin-filtering effect is lost for $\Delta_Z=0$, i.e.,
when the spin degeneracy is restored.

\section{Conclusion}
\label{Sec:Conclusion}
We have shown that spin filtering can be achieved by coupling the edge
states of the spin-split zeroth Landau level in graphene to a
superconductor. The spin-filtering effect can be switched on and off
by applying a (global) gate voltage that shifts the Fermi
energy. Unlike the charge conductance, the spin polarization is
independent of the edge termination. The device can be put in
different regimes by tuning the Zeeman energy independently of the gap
between the zeroth Landau level and the other Landau levels. This can
be achieved by applying an in-plane magnetic
field~\cite{giesbers09gap,kurganova11spin,chiappini15lifting}. The
spin filtering effect discussed here does not require the presence of
spin-orbit coupling and its experimental verification is within
the current technological capabilities.

\begin{acknowledgments}
This work was financially supported by the Swiss National Science
Foundation (SNSF) and the NCCR Quantum Science and Technology.
\end{acknowledgments}


\begin{thebibliography}{99}

\bibitem{novoselov07room} K.S. Novoselov, Z. Jiang, Y. Zhang,
S.V. Morozov, H.L. Stormer, U. Zeitler, J.C. Maan, G.S. Boebinger,
P. Kim, and A.K. Geim,
% {\it Room-Temperature Quantum Hall Effect in Graphene},
Science {\bf 315}, 1379 (2007).

\bibitem{heersche07bipolar} 
H.B. Heersche, P. Jarillo-Herrero, J.B. Oostinga, L.M.K. Vandersypen, and
A.F. Morpurgo, 
% {\it Bipolar supercurrent in graphene},
Nature {\bf 446}, 56 (2007).

\bibitem{jeong11observation} 
D. Jeong, J.-H. Choi, G.-H. Lee, S. Jo, Y.-J. Doh, and H.-J. Lee, 
% {\it Observation of supercurrent in PbIn-graphene-PbIn Josephson junction},
Phys. Rev. B {\bf 83}, 094503 (2011).

\bibitem{lee15ultimately}
G.-H. Lee, S. Kim, S.-H. Jhi, and H.-J. Lee, 
%{\it Ultimately short ballistic vertical graphene Josephson junctions}
Nat. Commun. {\bf 6}, 6181 (2015).

\bibitem{rickhaus12quantum}
P. Rickhaus, M. Weiss, L. Marot, and C. Sch\"{o}nenberger,
% {\it  Quantum Hall effect in graphene with superconducting electrodes},
Nano Lett. {\bf 12}, 1942 (2012).

\bibitem{park17propagation} 
G.-H. Park, M. Kim, K. Watanabe, T. Taniguchi, and H.-J. Lee, 
% {\it Propagation of superconducting coherence via chiral
% quantum-Hall edge channels},
Sci. Rep. {\bf 7}, 10953 (2017).

\bibitem{lee17inducing} 
G.-H. Lee, K.-F. Huang, D.K. Efetov, D.S. Wei, S. Hart, T. Taniguchi,
K. Watanabe, A. Yacoby, and P. Kim, 
%{ \it Inducing superconducting correlation in quantum Hall edge states},
 Nat. Phys. {\bf 13}, 693 (2017).

\bibitem{san-jose15majorana}
P. San-Jose, J.L. Lado, R. Aguado, F. Guinea, and J. Fern\'andez-Rossier,
% {\it Majorana Zero Modes in Graphene},
Phys. Rev. X {\bf 5}, 041042 (2015).

\bibitem{finocchiaro18topological}
F. Finocchiaro, F. Guinea, and P. San-Jose,
% {\it Topological $\pi$ Junctions from Crossed Andreev Reflection in the Quantum Hall Regime},
Phys. Rev. Lett. {\bf 120}, 116801 (2018).

\bibitem{peres06electronic} N.M.R. Peres, F. Guinea, and A.H. Castro Neto,
% {\it Electronic properties of disordered two-dimensional carbon},
Phys. Rev. B {\bf 73}, 125411 (2006).

\bibitem{brey06edge} L. Brey and H.A. Fertig,
% {\it Edge states and the quantized Hall effect in graphene},
Phys. Rev. B {\bf 73}, 195408 (2006).

\bibitem{abanin06spin} D.A. Abanin, P.A. Lee, and L.S. Levitov,
% {\it Spin-Filtered Edge States and Quantum Hall Effect in Graphene},
Phys. Rev. Lett. {\bf 96}, 176803 (2006).

\bibitem{akhmerov07detection}
A.R. Akhmerov and C.W.J. Beenakker,
% {\it Detection of Valley Polarization in Graphene by a
% Superconducting Contact},
Phys. Rev. Lett. {\bf 98}, 157003 (2007).

\bibitem{tworzydlo07valley} J. Tworzyd\l{}o, I. Snyman, A.R. Akhmerov,
and C.W.J. Beenakker
% {\it Valley-isospin dependence of the quantum Hall effect in a
% graphene p-n junction},
Phys. Rev. B {\bf 76}, 035411 (2007).

\bibitem{volkov12interaction}
A.V. Volkov, A.A. Shylau, and I.V. Zozoulenko, 
% {\it Lifting of the Landau level degeneracy in graphene devices in a
% tilted magnetic field}, 
Phys. Rev. B {\bf 86}, 155440 (2012).

\bibitem{beenakker06specular} C.W.J. Beenakker,
% {\it Specular Andreev Reflection in Graphene},
Phys. Rev. Lett. {\bf 97}, 067007 (2006).

\bibitem{greenbaum07pure} D. Greenbaum, S. Das, G. Schwiete, and P. G. Silvestrov,
% {\it Pure spin current in graphene normal-superconductor structures},
Phys. Rev. B {\bf 75}, 195437 (2007).

\bibitem{groth14kwant} C.W. Groth, M. Wimmer, A.R. Akhmerov, and X. Waintal,
% {\it {Kwant: a software package for quantum transport},
New J. Phys. {\bf 16}, 063065 (2014).

\bibitem{giesbers09gap}
A.J.M. Giesbers, L.A. Ponomarenko, K.S. Novoselov, A.K. Geim,
M.I. Katsnelson, J.C. Maan, and U. Zeitler,
% {\it Gap opening in the zeroth Landau level of graphene},
Phys. Rev. B {\bf 80}, 201403(R) (2009).

\bibitem{kurganova11spin}
E.V. Kurganova, H.J. van Elferen, A. McCollam, L.A. Ponomarenko,
K.S. Novoselov, A. Veligura, B.J. van Wees, J.C. Maan, and U. Zeitler,
% {\it Spin splitting in graphene studied by means of tilted
% magnetic-field experiments},
Phys. Rev. B {\bf 84}, 121407(R) (2011).

\bibitem{chiappini15lifting}
F. Chiappini, S. Wiedmann, K. Novoselov, A. Mishchenko, A.K. Geim,
J.C. Maan, and U. Zeitler, 
% {\it Lifting of the Landau level degeneracy in graphene devices in a
% tilted magnetic field}, 
Phys. Rev. B {\bf 92}, 201412(R) (2015).
\end{thebibliography}
\end{document}